\documentclass[12pt]{article} 
\usepackage{amsmath}
\usepackage{graphicx}

\title{Plasmonic modes of gold nano-particle arrays on thin gold films 
\footnote{A reviewed and edited version of this manuscript is published in 
Phys. Status Solidi RRL {\bf 4}(10), 256, 2010}}

\author{
  A. Hohenau\footnote{e-mail: andreas.hohenau@uni-graz.at}, J. R. Krenn \\
  Institute of Physics, Karl-Franzens University 
Graz, Austria
  }

\begin{document}

\maketitle

\section*{Abstract}
Regular arrays of metal nanoparticles on metal films have tuneable 
optical resonances that can be applied for surface enhanced Raman
scattering or biosensing. With the aim of developing more surface
selective geometries we investigate regular gold nanoparticle arrays on
25nm thick gold films, which allow to excite asymmetric surface
plasmon modes featuring a much better field confinement compared to the
symmetric modes used in conventional surface plasmon resonance
setups. By optical extinction spectroscopy we identify the plasmonic
modes sustained by our structures. Furthermore, the role of thermal
treatment of the metal structures is investigated, revealing the role of
modifications in the crystalline structure of gold on the optical
properties.

\section{Introduction}
Due to their spatially confined electron system, metal nanoparticles
exhibit collective electronic modes known as localized surface plasmons
(LSP)~\cite{Kreibig:MetalClusters}. These modes usually occur in the
visible spectral range and, if excited resonantly, give rise to
spatially highly confined (at the nm scale) optical field enhancement,
responsible for surface enhanced Raman scattering
(SERS)~\cite{Kneipp:Book} or surface enhanced
fluorescence~\cite{moskovits:85} effects. Optical field enhancement and
related surface enhanced effects can be further intensified by
electromagnetic interaction between adjacent
nanoparticles~\cite{Kneipp:Book}. In case of particle ensembles on
top of a metal surface, their mutual electromagnetic interaction is
partly mediated by surface surface plasmon polaritons
(SPP)~\cite{Raether:SurfacePlasmons}. This interaction was shown to
increase SERS efficiency~\cite{Felidj:PhysRevB:02}, however, a definite
assignment of the observed optical extinction peaks to distinct LSP/SPP
modes is still missing.

\begin{figure}
\begin{center}
\includegraphics[width=7.5cm]{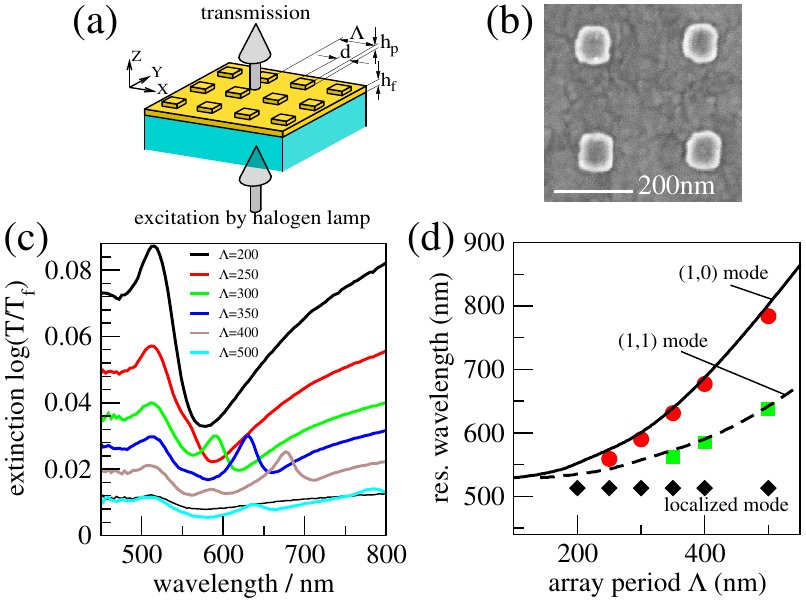}
\caption{(a) Sketch of the sample and measurement geometries. The 
samples are illuminated from the glass-substrate side, the transmitted light
is collected by a 2.5x, 0.075 numerical aperture objective and
analyzed by a spectrometer; (b) Exemplary SEM image; (c) Extinction 
spectra of particle arrays with varying array periods. The reference 
for calculating the extinction is taken on the gold film outside the
arrays.  The thin line depicts the extinction spectrum of a sample with
random  particle distribution (see text); (d) Resonance wavelengths of
the  extinction peaks vs. array period: experimental values from  (c)
(symbols) and calculated values (lines) of different grating orders.}
\label{fig:Sample}
\end{center}
\end{figure}

In this work, we investigate the optical modes of regular arrays of gold
nanoparticles fabricated by standard electron beam
lithography~\cite{Rai-Choudhury:Microlithography} directly on top of a
$25\:{\rm nm}$ thick gold film  (Fig.~\ref{fig:Sample}), with array 
periods of $200-500\:{\rm nm}$. The samples are
investigated by scanning electron microscopy (SEM) and optical
extinction spectroscopy. The investigation of similar systems was
reported recently~\cite{Hohenau:PhysRevB:06,Cesario:OptLett:05}, however
for different parameter regimes giving rise to different optical modes
than those investigated in the following. A systematic variation of the 
array periods allows us to identify the observed extinction peaks. Furthermore, we
assess the potential of such geometries for refractive index sensing.

\section{Extinction spectra and peak assignment}

Fig.~\ref{fig:Sample}(c) depicts the measured extinction spectra of
arrays of square gold nanoparticles [Fig.~\ref{fig:Sample}(a,b)] with
side length $d\sim 100\:{\rm nm}$, height $h_p=30\:{\rm nm}$ and array
periods $\Lambda=200-500\:{\rm nm}$ on a $h_f=25\:{\rm nm}$ thick gold
film. The overall size of the arrays is $100 \times 100\:{\rm \mu m}$.
For all array periods we observe one extinction peak at $\sim 520\:{\rm
nm}$ and, additionally, a second peak which shifts to larger wavelength
for larger array periods $\Lambda$. Additionally to these peaks, the
extinction spectra show a  rise towards longer wavelength. One can
assume that the shifting peaks are related to some coupling phenomenon
between the particles, while the period independent peak at $\sim
520\:{\rm nm}$ can be assigned to the excitation of a resonance
localized to the single particles. 

The spectral position of the peaks in the measured spectra
[Fig.~\ref{fig:Sample}(c)] are plotted in Fig.~\ref{fig:Sample}(d) as a
function of the array period (symbols). In addition, we plot the
calculated grating coupling resonances to antisymmetric SPP modes 
(`a-modes', defined by the symmetry of the tangential magnetic field with respect to 
the gold-film plane; the field maximum is at the gold-glass interface) in
different grating orders (lines). Due to too small array 
periods, symmetric SPP modes cannot be excited. The grating coupling 
resonances were calculated by finding
the spectral minima of the denominator of the Fresnel
coefficient~\cite{Burke:PhysRevB:86} of an unstructured gold film 
between glass and air. For the calculation the exciting light wave was
assumed to be laterally modulated  with a periodicity equal to the array
period $\Lambda$. We used the actual gold film thickness ($25\:{\rm
nm}$), the optical constants of gold from
Ref.~\cite{Johnson:PhysRevB:72} and a refractive index of $n=1.46$ for the
quartzglass substrate. 

As we find excellent agreement between experimental and calculated data
we conclude that grating coupling to a-mode SPPs is indeed observed. The
period independent peak at $\sim 520\:{\rm nm}$ can be assigned to a
combination of a vertically oriented dipole LSP resonance located at the
nanoparticles and scattering to high-energy SPP~\cite{Holland-Hall:1984}. 
To support this interpretation, we fabricated a sample with the same
average particle density as the $\Lambda=500\:{\rm nm}$ array but with
randomly distributed particles [thin black line in
Fig.~\ref{fig:Sample}(c)]. Indeed, in this case only the localized peak
and no grating related peaks are observed.

\begin{figure}
\begin{center}
\includegraphics[width=7.5cm]{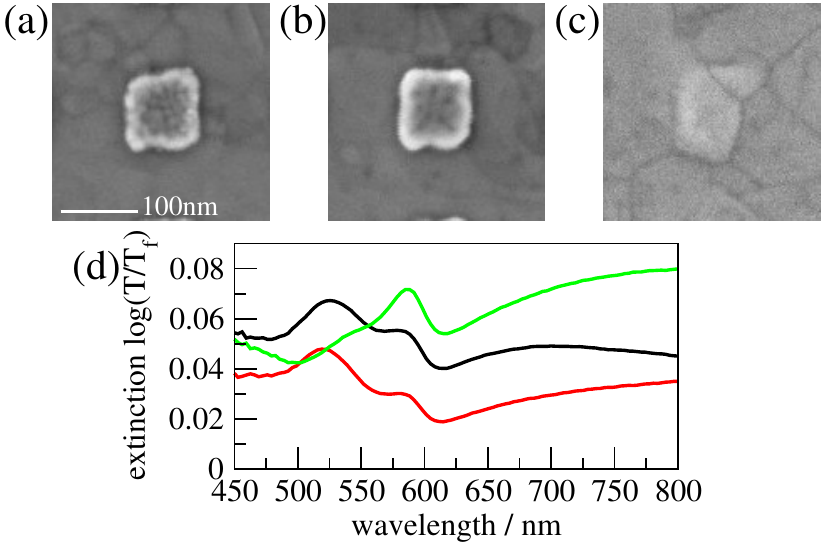}
\caption{(a) SEM images of a particle after fabrication;
(b) after annealing for 30 min at $175^{\circ}{\rm C}$; (c) after 4h at 
$200^{\circ}{\rm C}$; (d) Extinction spectra corresponding to (a) 
(black line), (b) (red line) and (c) (green line). 
\label{fig:annealing}}
\end{center}
\end{figure}

\section{Influence of thermal treatment}

By comparing the spectra in Fig.~\ref{fig:Sample}  with those published
by Felidj et al.~\cite{Felidj:PhysRevB:02}, one observes that the latter
show a broad extinction feature at $\sim 650-700\:{\rm
nm}$ instead of the monotonic rise above the grating coupling peaks as
observed in Fig.~\ref{fig:Sample}(c), although the geometry of the
samples was similar. This discrepancy could result from thermal
treatment of the gold film. When prepared by vacuum evaporation on glass
at room temperature, the gold thin films are built from crystallites
about $20\:{\rm nm}$ large. In the standard lithography process for
fabricating samples as used here (and in 
Ref.~\cite{Felidj:PhysRevB:02})
the gold film is covered by an electron
resist~\cite{Rai-Choudhury:Microlithography} which has to undergo an
annealing procedure ($175^{\circ}{\rm C}$ for 8  hours), leading to
a recrystallization of the gold~\cite{Aspens:PhysRevB:80}. In contrast,
the gold nanoparticles on top of the gold film are added later and hence
do not undergo thermal treatment. The differences in the surface
structure of gold film and particles due to the different crystallite
size is obvious in the SEM image in  Fig.~\ref{fig:annealing}(a). The
corresponding spectrum [black line in Fig.~\ref{fig:annealing}(d)] shows
indeed the same broad peak at  as observed in
Ref.~\cite{Felidj:PhysRevB:02}. Further experiments with
particles of different shapes and sizes (not shown), revealed no
significant change of the spectral position of the grating coupling
resonances or the appearance of the broad $\sim 650-700\:{\rm nm}$ peak.

If we now anneal the whole sample for 30 minutes at $175^{\circ}{\rm C}$,
recrystallization within the particles occurs while their shape is
maintained~\cite{Chen:NanoLetters:09} [Fig.~\ref{fig:annealing}(b)].
The spectra then show only a monotonic increase of the extinction
towards the red spectral range [red line in Fig.~\ref{fig:annealing}(d)
and spectra in Fig~\ref{fig:Sample}]. By further annealing (4h at
$200^{\circ}{\rm C}$), the particles loose their shape and melt into the
crystal grains of the  substrate [Fig.(\ref{fig:annealing}(c)]. In the
extinction spectrum, this leads to a disappearance of the localized peak
at $\sim 520\:{\rm nm}$ and an increase of the grating coupling peaks
[green line in Fig.~\ref{fig:annealing}(d)]. 
We assign  the observed differences upon thermal
treatment to changes in the particle's  dielectric function and shape,
in accordance with the finding reported in
Ref.~\cite{Chen:NanoLetters:09}.

\section{Influence of superstratum: potential for sensing applications}

Although featuring a field maximum at the gold-glass interface, the
a-mode SPP is of considerable strength at the gold-superstrate interface
for thin gold films (below $\sim 40\:{\rm nm}$). 
Importantly, the a-mode is bound strongly to the interface, giving
superior surface selectivity as compared to the conventional surface
plasmon resonance (SPR~\cite{Raether:SurfacePlasmons}) scheme, making
it potentially attractive for refractive index (RI) sensing. 

\begin{figure}
\begin{center}
\includegraphics[width=8cm]{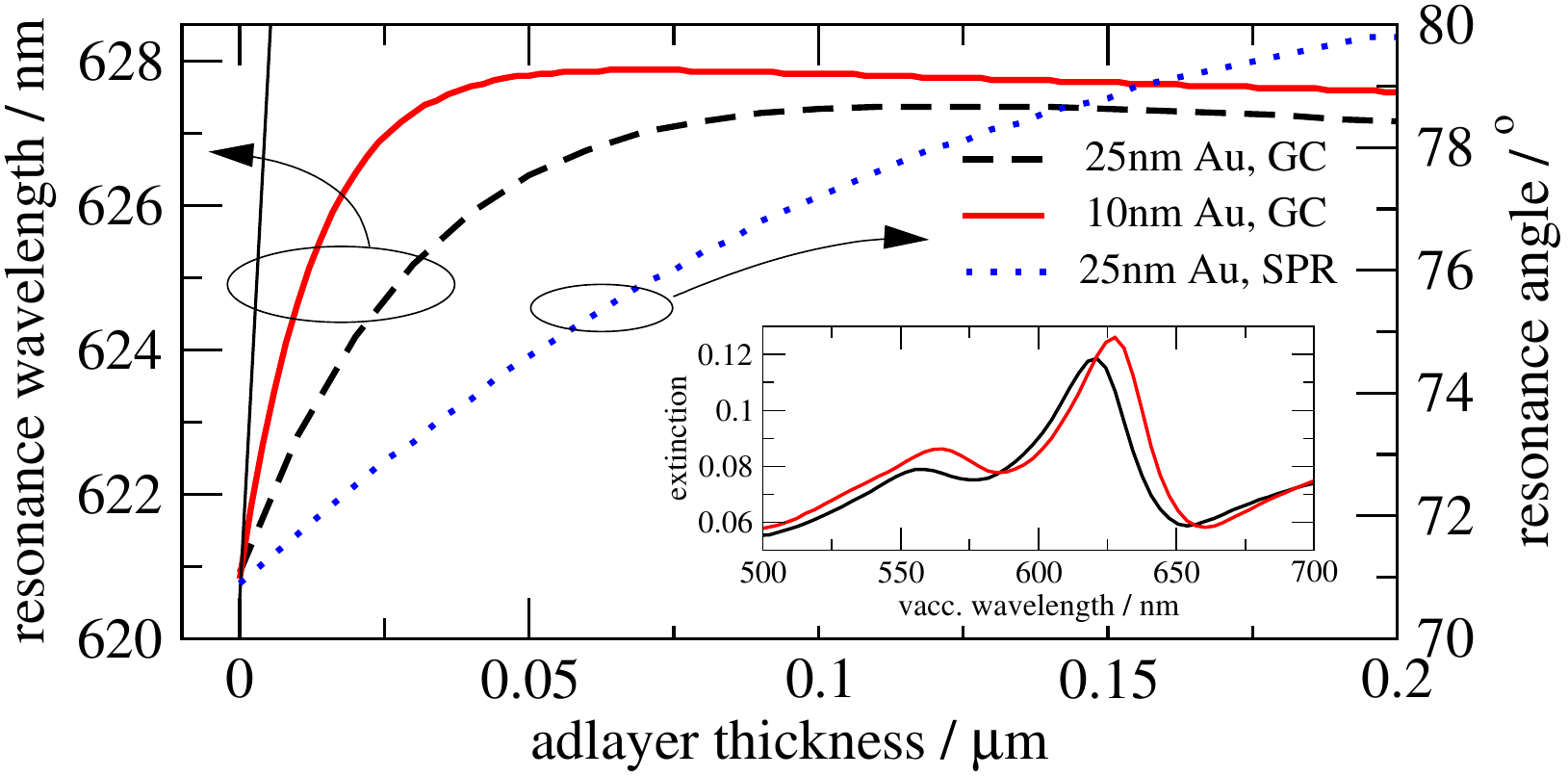}
\caption{\label{fig:sensing} Calculated comparison of the resonance
shift for a-mode grating coupling (GC) and SPR. The array periods are 
chosen as $\Lambda=151\:{\rm nm}$ ($\Lambda=285\:{\rm nm}$) for the
$10\:{\rm nm}$  ($25\:{\rm nm}$) film to get a resonance at $620\:{\rm
nm}$. The thin black line
shows the resonance wavelength of the SPR at a constant angle of
incidence of $71.8^{\circ}$. The inset depicts the measured extinction
spectra of an array with $\Lambda=350nm\:{\rm nm}$ and $h_f=25\:{\rm
nm}$ in air (black) and covered with $10\:{\rm  nm}$ thick $\rm SiO_2$ adlayer 
(red).}
\end{center}
\end{figure}

We illustrate this in Fig.~\ref{fig:sensing} by qualitatively comparing
the  shift of the calculated resonance positions of our arrays with that
of  the SPR angle. The investigated system consists of a standard glass substrate
($n=1.52$), a $10\:{\rm nm}$ ($25\:{\rm nm}$) thick Au film, an adlayer of varying thickness
($n=1.4$) and water ($n=1.33$) as the superstrate.  
While the slope of the spectral resonance shift of our grating coupling resonance of 
the $10\:{\rm nm}$ ($25\:{\rm nm}$) film is about a factor 4
(8) smaller than for the SPR resonance (thin black line in
Fig.~\ref{fig:sensing}), the sensing depth (defined as 63\% of the
maximum peak shift) is $\sim 100\:{\rm nm}$ for the SPR scheme but only
$\sim 13\:{\rm nm}$ ($\sim 26\:{\rm nm}$) for the grating coupling 
scheme, i.e., clearly less susceptible to bulk RI changes.

As a proof of principle, we experimentally measured the resonance shift 
of one of our arrays (after annealing for 4h at $200^{\circ}\:{\rm  C}$)
when adding a $10\:{\rm nm}$ thick $\rm SiO_2$ on the array-air side.
For the array with $\Lambda=350\:{\rm nm}$ we observe a resonance shift
of $9\:{\rm nm}$ of the (1,0) grating peak (inset of
Fig.~\ref{fig:sensing}). The estimated sensitivity of our current setup
is  $\sim 1\:{\rm nm}$ $\rm SiO_2$ in air. 

\section{Conclusion}
We demonstrated that the extinction spectra of regular arrays of gold
nanoparticles on $25\:{\rm nm}$ thick gold  films on glass substrate are
governed by extinction peaks related to grating coupling to a-mode SPP
with a field maximum at the glass-gold interface, and by a resonance
localized to  the single particles at $\sim 520\:{\rm nm}$. By annealing
of the array the particles fuse with the crystal structure of  the
substrate, the localized resonance disappears and the grating coupling
is enhanced. Such a geometry was used to demonstrate the feasibility of
sensing a thin dielectric layer on top of the array by monitoring the
resonance wavelength of the grating coupling peaks. The sensitivity is
less than for a usual SPR sensor, the surface selectivity is however
increased by a factor of 4-8 due to the better spatial confinement of
the SPP near fields. As an additional advantage we note that due to the
grating coupling scheme there is no need for the substrate refractive
index to be larger than the superstrate index.

\section*{Acknowledgement}
We acknowledge support from the Austrian Science Foundation FWF under 
grant Nr. P21235-N20.

\bibliographystyle{unsrt}

\end{document}